\documentclass[aps,prb,twocolumn]{revtex4}

\usepackage{graphics,amsmath}

\usepackage{graphicx}

\begin{document}

\title{Signatures of polaronic charge ordering in optical and dc
    conductivity using dynamical mean field theory.}

\author{S. Ciuchi}
\affiliation{SMC Research Center INFM-CNR, CNISM and
Dipartimento di Fisica\\
Universit\`a dell'Aquila,
via Vetoio, I-67010 Coppito-L'Aquila, Italy}

\author{S. Fratini}
\thanks{On leave of absence at the ICMM-CSIC, Madrid, Spain.}

\affiliation{Institut NEEL, CNRS \& Universit\'e Joseph Fourier\\
BP 166, F-38042 Grenoble Cedex 9, France}

\begin{abstract}

We apply dynamical mean field theory to study a prototypical model that
describes charge ordering in the presence of both electron-lattice
interactions and intersite electrostatic repulsion between electrons.
We calculate the optical and d.c. conductivity, and derive approximate
formulas valid  in the limiting electron-lattice coupling regimes.
In the weak coupling regime, we recover the usual behavior of charge
density waves,  characterized  by a transfer of spectral weight due to
the opening of a gap in the excitation spectrum. In the opposite
limit of very strong electron-lattice coupling, instead,
the charge ordering transition is signaled by a global enhancement of the
optical absorption, with no appreciable spectral weight transfer. 
Such behavior is related to the progressive suppression 
of thermally  activated charge defects taking place below the critical
temperature.  
At  intermediate values of the coupling within the polaronic regime, 
a complex behavior is obtained 
where  both mechanisms of transfer and enhancement of spectral weight
coexist.

\end{abstract}

\date{\today}


\maketitle

\section{Introduction}

As opposed to  conventional 
charge density waves, \cite{Gruner}
that are well understood in terms of  lattice-driven instabilities of the Fermi
surface in metallic systems,
there is no unified description of the charge ordering (CO) transitions 
observed in strongly interacting systems, or ``bad metals''. 
Typical examples of such systems are transition-metal oxides, which
%These systems 
are invariably
characterized by a complex interplay between several microscopic
interactions, involving the charge, spin, orbital and lattice degrees of
freedom. This complexity seems to 
preclude the identification of a simple, common,  charge
ordering mechanism. Still, electron-lattice interactions
are always present to some extent, and 
often play a dominant role in driving the 
CO transition  in these systems. 
Charge ordering phenomena
involving a strong electron-lattice coupling 
have been found in 
manganites,\cite{Dessau,Adams,Vasiliu} nickelates,\cite{Katsufuji,Calvani} 
layered cobaltates,\cite{Bernhard,Wang}  
magnetite, \cite{Degiorgi,Park,Schrupp} 
vanadates, \cite{Presura,Baldassarre} oxoborates,\cite{Attfield} as
well as in other inorganic low-dimensional systems.\cite{Perfetti1,Perfetti2} 
The lattice degrees of freedom 
could also be relevant in the charge ordered phases 
of the so-called ``telephone-number'' ladder compounds,\cite{Vuletic}
as well as in two-dimensional organic salts,\cite{Drichko} 
although in those cases most theoretical interpretations up to now have focused
on models with purely electronic interactions.

In this work we focus on a minimal model which describes 
charge ordering in the presence of electron-lattice 
interactions, with particular attention to the polaronic regime
obtained at strong coupling. 
The model consists of electrons on a bipartite
lattice, at a commensurate concentration of one electron on every two
sites, interacting locally with dispersionless lattice vibrations. 
The electrons also interact mutually via an intersite electrostatic repulsion,
which can be thought of as the screened part of the long-ranged
Coulomb potential.
While this oversimplified model only retains part of the 
complex physics involved in real 
systems, its solution 
can be helpful to clarify certain aspects of the charge
ordering phenomena that are common to  
systems with strong electron-lattice interactions, 
and that can in principle be identified experimentally.
To be  specific, 
we solve the model by  performing the following approximations:

(i) The magnetic degrees of freedom are taken out of the game by
resorting to spinless electrons. This enforces
locally the constraint of no double occupancy characteristic of the  limit of
strong  ``Hubbard'' (on-site) repulsion. 
It is appropriate for our purposes, as long as we 
do not aim at describing the effects of electronic correlations on the
low-energy physics, such as the existence of a 
quasi-particle peak or the induced magnetic exchange.
In principle, this approximation is viable as long as 
the magnetic and charge degrees of freedom 
are governed by different energy scales. In such case,  the magnetic
interactions are not expected to have a strong influence on the charge
ordering pattern (although exceptions do exist
\cite{Bulla99,Bulla04,Daghofer}).  
This separation of energy scales is realized for instance 
in magnetite (Fe$_3$O$_4$),\cite{Walz} 
where the CO transition occurs within a
ferromagnetic phase that preexists at a much higher temperature. 
More examples can be found in systems (such as V$_3$O$_5$
\cite{Baldassarre}) where the CO transition takes 
place within a paramagnetic phase, the magnetic order setting in at a
much lower temperature.  The
spinless approximation is also relevant to half-metals such as the
colossal magnetoresistance manganites.

(ii) We  consider a single electronic
band, which rules out the orbital degrees of freedom and
their possible ordering, that is known to play an important role in
specific compounds.

(iii) The lattice vibrations are assumed adiabatic,
i.e. their characteristic energy is much smaller than the electronic
bandwidth, which is typically the case in transition-metal oxides. 
The adiabatic approximation is enforced here 
by treating the lattice degrees of freedom
as static variables with a given (thermal) statistical distribution.
As a consequence, the low-energy spectral features
that derive from the quantum nature of the phonons are lost. These
would appear  at energies comparable to, or below the phonon energies. 
Such features are anyhow of minor quantitative
importance in the polaronic regime of interest here, where most of 
the spectral weight is moved to higher energies, of the order of the
polaron binding  energy. The same argument justifies
the neglect of low-energy phenomena originating from electronic
correlations as discussed
at point (i) above, in all cases 
where the polarons set the dominant energy scale.

(iv) We apply single-site Dynamical Mean Field Theory (DMFT), 
appropriately adapted\cite{Ciuk99,Bulla99,Hassan07,Matveev08}  
to account for the charge unbalance between
neighboring sites in the charge ordered state.
This approach is known to deal
very effectively with local interaction mechanisms, regardless of their
strength. It is therefore
expected to give an accurate description of the electron-lattice
physics. As was shown in Ref. \onlinecite{Ciuk99}, this approach
correctly describes  the  crossover between the weak coupling charge
density wave regime and strong-coupling polaronic regime, 
and allows to shed light on the 
role played by defects of the lattice polarization 
at the ordering transition.  
A lesser accuracy is achieved in treating the inter-site Coulomb 
repulsion, since non-local interactions reduce to the mean-field 
(Hartree) level in single-site DMFT. 
This is however not a primary limitation\cite{Bulla99,Bulla04} 
in the regime we are mainly interested in, where 
the physics is dominated by the electron-lattice interaction.

The main result of this work is the identification of qualitatively 
different behaviors of the electrodynamic response in the  
two fundamentally distinct regimes of charge ordering, governed by
the strength of the electron-lattice interaction. In the weak
coupling regime, the electrodynamic response is dominated by the
opening of an energy gap in the excitation spectrum of an
otherwise  metallic system. At the ordering transition, 
the optical spectral weight
is transferred from the Drude peak to the region above the gap, and
the d.c. conductivity acquires a semiconducting character. 
On the contrary, in the strong
coupling limit, charge localization is already present in the
normal phase, due to the formation of self-localized polarons.
Correspondingly, the low energy spectrum is already strongly depleted
by a polaronic ``pseudo-gap'',  and the genuine gap opening occurring at the
transition does not give rise to any appreciable depletion or shift of
the low frequency spectral weight.
Instead, the ordering of polarons results in
a global enhancement of the optical spectral weight, which is
opposite to what is expected from a conventional charge density wave. 
The two scenarios described above coexist in the polaronic regime
at moderate values of the electron-lattice coupling. This leads to a
complex behavior presenting both a transfer and an enhancement
of the spectral weight, resulting in
a non-monotonic temperature dependence of the
total optical  weight within the CO phase.
The inclusion of a direct intersite Coulomb
interaction extends this complex intermediate region to larger
electron-lattice interaction strengths.

The solution for the one-particle spectral function  
has already been published in Ref. \onlinecite{Ciuk99}, in a model
with electron-lattice interactions alone.
Here we extend that treatment to calculate the optical and
d.c. conductivity throughout the parameter space,  including 
an intersite Coulomb repulsion between electrons. 
We also derive analytical expressions
for these quantities that are valid in the polaronic regime,
elucidating the role played by  charge defects in both optical
absorption and charge transport.
It can be noted that an analogous model, without the Coulomb interaction
term, was studied variationally in Ref. \onlinecite{Perroni}. There, however, 
the physics of defects  was not addressed.

The paper is organized as follows.
The general formalism for treating the charge ordered, 
broken-symmetry phase within
the DMFT is presented in Sec. II. The equations are solved numerically
in Sec. III. The optical and d.c. conductivity are calculated in
Sec. IV and V respectively, making use of the Kubo formula. The
results are  discussed in Sec. VI.

\section{Model and self-consistent solution}
\label{sec:sol}

We consider the following Holstein-Coulomb Hamiltonian:
\begin{eqnarray}
  H&=&-\frac{t}{\sqrt{z}}\sum_{<ij>}(c^+_i c_j +
  h.c.)+\frac{1}{2}k X_i^2  \label{eq:H} \\
 & &+ g\sum_i(c^+_ic_i-n)X_i +\frac{V}{2z}\sum_{<ij>}(n_i-n)(n_j-n)
\nonumber
\end{eqnarray}
where the operator $c^+_i (c_i)$ creates (destroys) 
a spinless electron at site $i$, $n$
is the average electron density per site and
$X_i$ 
are the displacements 
of local oscillator modes, that are coupled to the local electron
density via the parameter $g$. In addition to this electron-phonon
interaction
{\it \`a la Holstein}, the electrons interact mutually via a nearest
neighbor repulsion term $V$ of Coulomb origin.  
The scaling of the repulsive term is chosen in
such a way  that the energy cost
for a deviation from perfect charge ordering (i.e. flipping an
occupied and an empty site) equals $V$.
The scaling of the kinetic term $t/\sqrt{z}$ yields
a finite free-electron bandwidth in the limit of
infinite connectivity (number of nearest neighbors $z\to \infty$).
We shall
consider for simplicity  a semi-elliptical density of states (DOS) of
half-width $D=2t$, corresponding to a Bethe lattice of infinite connectivity.
Nevertheless, the present
calculation scheme can be easily generalized to other model DOS, to
include more realistic band structures as 
obtained for example from ab-initio calculations of specific systems.
We shall specialize to the commensurate concentration $n=1/2$,
which is equivalent to a quarter-filled band in the case of electrons with
spin. In the present spinless case,  the chemical potential is fixed
at $\mu=0$ from particle-hole symmetry.

An order parameter for the CO transition can be defined as
the charge disproportionation
\begin{equation}
  \label{eq:po}
  \Delta n=n_A-n_B,
\end{equation}
which varies between $0$ and $1$, the average density being given by
$n=(n_A+n_B)/2=1/2$ (here A and B label respectively the
charge-rich and charge-poor  sublattices).
 The formalism needed to treat the electron-phonon interaction
was set up in Ref. \onlinecite{Ciuk99} by  mapping the
original lattice problem onto a
pair of coupled impurity models. In 
 the limit of infinite connectivity,   any additional
non-local interaction term self-averages to its mean-field
(Hartree) value. 
Therefore, the extra Coulomb
term can be straightforwardly included  via a site dependent chemical potential
shift in the Weiss fields, i. e. the local fields that take into
account the effects of the electron itinerancy. 
On the Bethe lattice, the new Weiss fields  read: 
\begin{eqnarray}
  \label{eq:Weiss}
\left( G_0^{-1} \right)^A=i\omega_n+\frac{V}{2} \Delta n
-\frac{D^2}{4} G^{BB}   \\
\left( G_0^{-1} \right)^B=i\omega_n-\frac{V}{2} \Delta n
-\frac{D^2}{4} G^{AA}.\label{eq:Weiss2} 
\end{eqnarray}
It can be observed by direct inspection of Eqs. 
(\ref{eq:Weiss}-\ref{eq:Weiss2}) that the Coulomb interaction does not
affect the properties of the system in the normal phase (where  $\Delta n=0$), 
which is a drawback of the present approximation.

The above system of equations is closed self-consistently 
by expressing the site-diagonal propagators $G^{AA}$ and $G^{BB}$ in
terms of  the self-energy arising from the local interaction with the
static phonon field:
\begin{equation}
  \label{eq:self-con}
  G^{AA}=\int dX \frac{P_A(X)}{(G_0^{-1})^A-g X},
\end{equation}
and similarly for $G^{BB}$. Here $P_{A}(X)$ and $P_B(X)$ are the
probability distribution functions (PDF)
 for the phonon displacements on the A and B
sublattices, which can be obtained by tracing out the electronic
degrees of freedom as\cite{Ciuk99}
\begin{equation}
\label{eq:PDF}
P_{A}(X) \propto e^{-\beta k X^2/2} \prod_n  
\left[(G_0^{-1})^A-gX\right] e^{i\omega_n 0^+}.
\end{equation}
The numerical solution is achieved by successive  iterations of the
above equations, starting from a given ansatz for the sublattice
propagators $G^{AA}$ and $G^{BB}$. In the following we shall 
take the half-bandwidth $D$ as the unit of energy.

\begin{figure}
  \centering
  \includegraphics[width=80mm]{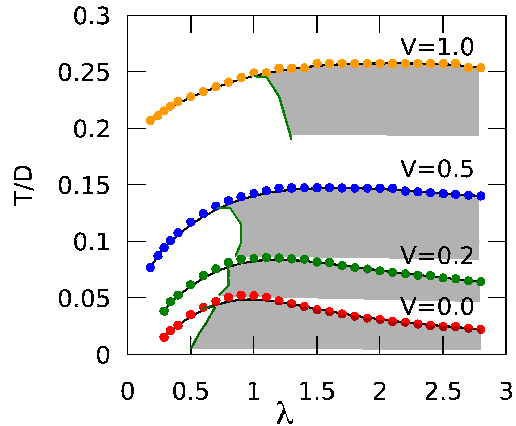} 
  \caption{Critical temperature delimiting the CO phase 
  as a function of $\lambda=g^2/2kD$, for different values of $V/D$. 
  Shaded areas below $T_c$ mark polaronic
  regions where defects are present. The 
  existence of defects is tracked by looking
  at metastable minima of the adiabatic potentials $V_{A,B}=-T\log(P_{A,B})$.}
  \label{fig:Tc}
\end{figure}

We now briefly analyze the phase diagram obtained from the
DMFT solution of the model Eq. (\ref{eq:H}).
Fig. \ref{fig:Tc} shows the critical temperature for the charge
ordering transition as a function of  the electron-phonon coupling parameter
$\lambda=g^2/2kD$, 
defined as the polaron energy $E_P=g^2/2k$  
in units of the half-bandwidth. At $V=0$, we recover the
bell-shaped result obtained in Ref. \onlinecite{Ciuk99}.  
Such bell shape originates from the different mechanisms that drive the
charge ordering in the 
two limiting regimes of small and large $\lambda$. 
At weak coupling,  the
critical temperature monotonically {\it increases} with the
electron-phonon coupling, i.e. the same qualitative trend 
predicted by the usual BCS theory. \cite{note-BCS}
In this regime, the charge disproportionation and the 
long-range order take place simultaneously at $T_c$. 
At strong coupling instead, due
to polaron formation, a {\it local} charge segregation corresponding
to the localization of the electrons on randomly distributed sites occurs at a
temperature $T\approx E_P$, much larger than ordering temperature
$T_c$ itself. The actual critical temperature $T_c$ 
marks the onset of spatial ordering
of such   randomly distributed polarons. It is proportional to
the ``charge superexchange''  $J=D/(4\lambda)$ and
therefore {\it decreases} as the 
electron-phonon coupling increases.  
The asymptotic strong coupling
expression for
the critical temperature is $T_c=D/(16 \lambda)$. From the above discussion we see that
the maximum of $T_c$, obtained at $\lambda\approx 1$ for $V=0$, 
signals the crossover 
between the weak-coupling and the polaronic behavior.

The ordered phase below
$T_c$ is also qualitatively different depending on the value of the 
electron-phonon coupling. In particular, in the 
strong coupling regime,  the preexisting polarons progressively order
upon lowering the temperature, and only attain a perfect alternate order at
$T=0$. At any finite temperature,
a certain number of defects exist within the CO phase,
in the form of charges localized on the ``wrong''
sublattice. The existence of such thermally-induced 
defects is 
intimately related to 
the existence of local lattice distortions in the normal phase,
i.e. polaron formation.\cite{Ciuk99}
We can therefore identify the existence of a
{\it polaronic} CO phase, as opposed to the weak-coupling
charge density wave, as the phase in which these defects are present.

The defect phase can be quantitatively characterized by analyzing the
sublattice PDF's, which acquire a bimodal structure for
sufficiently  large $\lambda$. 
More efficiently, one can  look for the existence of
two non-degenerate minima in  the sublattice adiabatic potentials defined 
as $V_{A,B}=-T\log{P_{A,B}}$. 
The result of this procedure is illustrated by the shaded
areas in the phase diagram of Fig.  \ref{fig:Tc}. We see that
the defect phase emerges right 
at the polaron crossover, i.e. in the region below
the maximum of $T_c$ vs $\lambda$.

Including a direct intersite repulsion 
clearly favors both types of  charge ordering,
as it increases $T_c$ for all values of $\lambda$ 
(compare the different curves in Fig. \ref{fig:Tc}). 
Still, the evolution of the phase diagram  with $V$ suggests 
a non-trivial interplay between the direct Coulomb repulsion
and the polaronic physics,
as a finite $V$  shifts the polaron crossover 
(as well as the boundary of the defect phase) 
towards larger values of $\lambda$.
It can be argued that, by treating the intersite Coulomb
interaction beyond mean-field, charge fluctuations 
of the same nature of the lattice defects evidenced above 
could start playing a role.\cite{Frat1D}
For this reason, in the following we shall
restrict to the regime of small to moderate $V$, 
where the dominant physics is set by the electron-phonon interactions, 
therefore suppressing the fluctuations of the Coulomb interaction 
term.

\begin{figure}
  \centering
  \includegraphics[width=70mm]{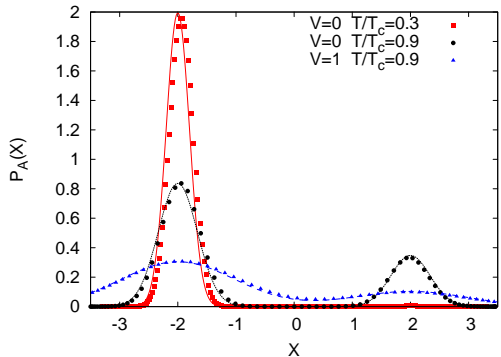}
  \caption{Sublattice phonon PDF at $\lambda=2$. The different curves
    are for $V=0$
    ($T/T_c=0.3$ and $0.9$) and $V=1$ ($T/T_c=0.9$). 
Points correspond to  the DMFT self-consistent solution,   
lines represent the strong coupling theory
 Eq. (\ref{eq:PXco}). The locus and shape of the peaks does not depend
 explicitly on $V$.}   
  \label{fig:defects}
\end{figure}

The sublattice PDF $P_A(X)$ evaluated at $\lambda=2$, well inside the
polaronic regime, is shown in Fig. \ref{fig:defects}. 
Defects in the $A$ sublattice give rise to a
minority peak of opposite ``polarization'' with respect 
to the  preferred one.
The area of the minority peak 
can be used to 
define  the number of defects ($n_d$) in the CO phase.
In the strong coupling regime, it  tends to $1/2$ at $T=T_c$, where
the two peaks become equivalent 
(polarons are randomly distributed over the two sublattices), while
it vanishes exponentially at $T=0$, 
when perfect charge ordering is achieved. This occurs because,
as the temperature is reduced, the population of the metastable
minimum of the potentials $V_{A,B}$ is progressively depleted, 
leading to an exponential
reduction of $n_d$. It should be noted that the possibility of 
strictly $n_d=0$ 
is related to our classical treatment of the lattice degrees of
freedom, which breaks down at temperatures much smaller
than  the characteristic phonon energies. 
Properly including phonon quantum fluctuations\cite{Blawid2}
would lead to the appearance of charge defects even in the zero
temperature limit.

In the limit 
$\lambda\to \infty$, the number of defects is related to the order parameter
through
\begin{equation}
\label{eq:defects}
\Delta n=1-2n_d.
\end{equation}
At finite $\lambda$ in the polaronic phase ($\lambda\gtrsim 1$), 
the above Eq. (\ref{eq:defects})  has to be replaced by 
$\Delta n=(1-2n_d)[1-1/(8\lambda^2)+\ldots]$, which properly accounts for the 
fact that the charge disproportionation $\Delta n$ 
does not strictly tend to $1$ even when the
polarons are perfectly ordered ($n_d=0$).
This is because at finite $\lambda$ 
the electronic wavefunction is not fully localized and
acquires a finite extension  on the
neighboring sites, which necessarily pertain to the minority sublattice.

\begin{figure}
  \centering
 \includegraphics[width=80mm]{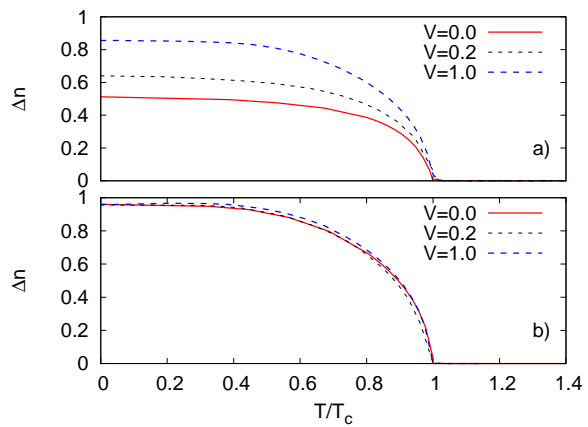}
  \caption{
Evolution of the charge disproportionation  for $\lambda=0.4$ (a) and for
   $\lambda=2.0$ (b), at different values of $V$.}
  \label{fig:po}
\end{figure}

The evolution of the charge disproportionation
with $T$ in different regions of the phase diagram
is reported in Fig. \ref{fig:po}. In the polaronic regime, 
$\Delta n$ is a universal function of
$T/T_c$, independent of $V$, as shown in Fig. \ref{fig:po}.b. 
This means in particular  that  the number
of defects $n_d$ is a  function of $T/T_c$ alone, and $V$ only
enters implicitly through the determination of $T_c$. 
Interestingly, the gap ratio  $\Delta_{T=0}/T_c$ can become
very large in the polaronic regime, due to the preexisting
pseudogap  $\Delta \propto E_P\gg T_c$. \cite{Ciuk99} 
In the opposite weak electron-phonon coupling limit,  we recover a 
BCS-like behavior for $\Delta n$. In that case the gap
is given by 
$\Delta=(\lambda D+V/2) \Delta n $.
Its value at $T=0$ satisfies the mean-field ratio 
$\Delta_{T=0}/T_c=3.54$. 
It is worth noting that this limit is only reached asymptotically as
$\lambda\rightarrow 0$. For small but finite $\lambda$, $T_c$ is dramatically
reduced by thermally induced lattice fluctuations, \cite{Ciuk99,Blawid} while 
$\Delta_{T=0}$ is not. Therefore the gap/$T_c$ ratio rapidly increases with
$\lambda$ even well outside the polaronic region.

\section{Optical conductivity}

\subsection{DMFT formulation }
We now derive the Kubo formula for the optical
conductivity in the CO phase.
To this aim we develop a general formalism
which makes explicit use of translational invariance as in the
hypercubic lattice,  following the lines of
Ref. \onlinecite{Kotliar}. 
In this framework it is possible to write 
expressions for the electron propagators in $k$-space. 
This is achieved through a canonical transformation that defines new 
electron creation operators
$c^{A,B}_{\bf k}=(c_{\bf k}\pm c_{{\bf k}+{\bf
    Q}})/\sqrt{2}$, where  
${\bf Q}=(\pi,\pi,\cdots)$ is the instability wave vector in any dimensions  
and ${\bf k}$ spans the reduced Brillouin zone (RBZ). 
With the above transformation, the tight binding term
becomes $\sum_{k}^{RBZ} \epsilon_{\bf k} (c^{\dagger A}_{\bf
    k}c^{B}_{\bf k}+c^{\dagger B}_{\bf k}c^{A}_{\bf k} )$, 
where 
$\epsilon_{\bf k}$ is the original non-interacting band dispersion.
The fully interacting
fermion propagators on the bipartite lattice can be defined as the
matrix elements
\begin{equation}
  G^{\alpha\beta}_{\bf k}=-i\langle T c^\alpha_{\bf k}(t) c^\beta_{\bf
  k}(0) \rangle \;\;\;\; (\alpha,\beta=A,B)
\label{eq:Gdef}
\end{equation} 
of the $2\times 2$ matrix $\hat{G}_{\bf k}$. The corresponding
spectral functions are given by $\hat{\rho}({\bf k},\omega)=-Im
\hat{G}_{\bf k}(\omega)/\pi$.   
Once the local self-energies on the two sublattices are known from
the solution outlined in Sec. \ref{sec:sol}, the
Green's functions of Eq. (\ref{eq:Gdef}) are obtained by inverting the matrix 
\begin{equation} \label{eq:invert}
\hat{G}_{\bf k}^{-1}=
\left(
\begin{array}{cc}
z^A & -\epsilon_{\bf k} \\
-\epsilon_{\bf k} &z^B 
\end{array}
\right)
\end{equation}
with $z^A=\omega+i \delta +\mu -\Sigma_A(\omega)$, and similarly for $z^B$.

In the tight binding model of Eq. (\ref{eq:H}), 
the current operator along a given (say $x$)
direction reads
\begin{equation}
  \label{eq:current}
  { J_x} = -i\frac{t}{\sqrt{z}}
\sum_{i,\hat\delta} \delta_x c^\dagger_{i+{\hat{\delta}}} c_i, 
\end{equation}
where the sum extends over all sites $i$ of the lattice and their $z$
nearest neighbors,  identified by the vectors ${\hat\delta}$. 
Transforming to the sublattice operators, 
the current operator can be expressed as: 
\begin{equation}
  { J_x}= -i \sum_{{\bf k} \in  RBZ} v_{\bf k} (c^{\dagger A}_{\bf
    k}c^{B}_{\bf k}+c^{\dagger B}_{\bf k}c^{A}_{\bf k} ).
\end{equation}
In a hypercubic lattice, only
the two neighbors along the $x$ direction contribute to the above sum,
and  the corresponding
current vertex is $v_{\bf k}=2\frac{t}{\sqrt{z}} \sin k_x$.

In the context of DMFT, due to the vanishing of  vertex 
corrections  \cite{Khurana,Schweitzer,Jarrell} 
the current-current correlation function can be expressed exactly in terms of
single particle Green's functions. This simplification still holds 
in the broken-symmetry phase because the ${\bf k}\to -{\bf k}$ symmetry is preserved even in 
the reduced Brillouin zone. 
Reminding the definitions Eq.(\ref{eq:Gdef}), 
we can write
the current-current correlation function as
\begin{eqnarray}
\langle T J_x (t) J_x (0) \rangle = \sum_{{\bf k} \in  RBZ} \;
v_{\bf k}^2 \; Tr \left [ \tau_x \hat G_{{\bf k}}(t)\tau_x \hat G_{{\bf
      k}}(-t)  \right]\label{eq:corr-corr}
 \end{eqnarray}
where we have made use of  the Pauli matrix $\tau_x$ 
to obtain a compact expression. The trace is performed in the
sublattice indices $A,B$.

The optical conductivity follows from the Kubo formula, 
upon transforming the current-current
correlation function to the frequency domain:  
\begin{eqnarray}
& & \sigma(\omega) =\sigma_0  \int_{RBZ} d\epsilon
N(\epsilon)\phi(\epsilon) \times \label{OC}\\
\times  & &\int_{-\infty}^{\infty}  d\nu\;  
{Tr} \left [ \tau_x \hat\rho (\epsilon, \omega+\nu) 
\tau_x \hat\rho (\epsilon, \nu) \right]
\; \frac{f(\nu)-f(\nu+\omega)}{\omega}.
\nonumber
\end{eqnarray}
In the above equation, the constant
$\sigma_0 =\pi e^2a^2 /\hbar v$ carries the dimensions
of conductivity, $a$ being the lattice spacing, $v$ the
volume of the unit cell, $f(\nu)=[1+e^{\beta (\nu-\mu)}]^{-1}$ 
the Fermi function and $\hat\rho$  the spectral functions associated to 
sublattice propagators $\hat G$. 
Taking advantage of the local nature ($k$-independence) of the self-energy, 
the  sum over momenta in Eq. (\ref{eq:corr-corr}) has been replaced by an
integration over energies, weighted by the DOS 
$N(\epsilon)$ and the current
vertex $\phi(\epsilon)$ of the non-interacting lattice. 
Use has also been
made of the fact that
$v_{\bf k}^2$ is invariant under the transformation 
${\bf k}\to  {\bf k}+{\bf Q}$. 
The integration over the band dispersion $\epsilon$ in the
current-current correlation function can be
performed analytically, leading to:
\begin{equation}
  \label{eq:sigma-bub}
   \sigma(\omega) = \sigma_0\int_{-\infty}^{\infty}  d\nu\;  
B(\nu+\omega,\nu) \; \frac{f(\nu+\omega)-f(\nu)}{\omega}.
\end{equation}
The function $B$ is defined as: 
\begin{widetext}
\begin{equation}
B(\nu+\omega,\nu)=-\frac{1}{4\pi} Re \left\lbrace \chi
(z^A_1,z^B_1;z^A_2,z^B_2)-\chi (z^A_1,z^B_1;z^{A*}_2,z^{B*}_2)\right\rbrace
\end{equation}
where
\begin{equation}
  \chi (z^A_1,z^B_1;z^A_2,z^B_2)=\frac{2}{\xi^2_1-\xi^2_2}
 \left\lbrace \frac{\mathcal{K}(\xi_2)}{\xi_2} 
          [z_1^Bz_2^A+z_1^Az_2^B+2\xi^2_2]  -
          \frac{\mathcal{K}(\xi_1)}{\xi_1} 
          [z_1^Bz_2^A+z_1^Az_2^B+2\xi^2_1] 
\right\rbrace  
\end{equation}
\end{widetext}
and
\begin{eqnarray}
  z_1^{\alpha}&=&\omega+\nu+i\delta -\Sigma_{\alpha}(\omega+\nu)\\
  z_2^{\alpha}&=&\nu+i\delta -\Sigma_\alpha(\nu)\\
  \xi_1&=&\sqrt{z^A_1z^B_1}\\
  \xi_2&=&\sqrt{z^A_2z^B_2}
\end{eqnarray}
with $\alpha=A,B$ the sublattice index. The function 
 $\mathcal{K}$ is the Hilbert transform of the product
$N(\epsilon) \phi(\epsilon)$, that can be evaluated  analytically 
in the case of a semi-elliptical DOS. For this we take  the following
form of the current vertex $\phi(\epsilon)=(D^2-\epsilon^2)/3$. This is  
chosen in such a way as to fulfill the f-sum rule
\cite{Freericks,Millis,PRBoptcond} 
which relates the total optical spectral weight 
\begin{equation}
W=\int_0^{\infty} d\omega \sigma(\omega).
\label{eq:optcond_sw}
\end{equation}
to the kinetic energy $K$ of the interacting system. 
The demonstration of the f-sum rule in the broken-symmetry phase
is explicitly carried out in Appendix C.

In the following Sections we shall analyze separately the results in 
the weak electron-phonon coupling regime $\lambda\lesssim 1$, in the
extreme strong coupling limit $\lambda\gg 1$, and in the most interesting
polaronic regime $\lambda\gtrsim 1$.

\subsection{Weak coupling regime,  $\lambda\lesssim 1$}
\label{weak}

\begin{figure}[htbp]
\centering
\includegraphics[scale=0.7]{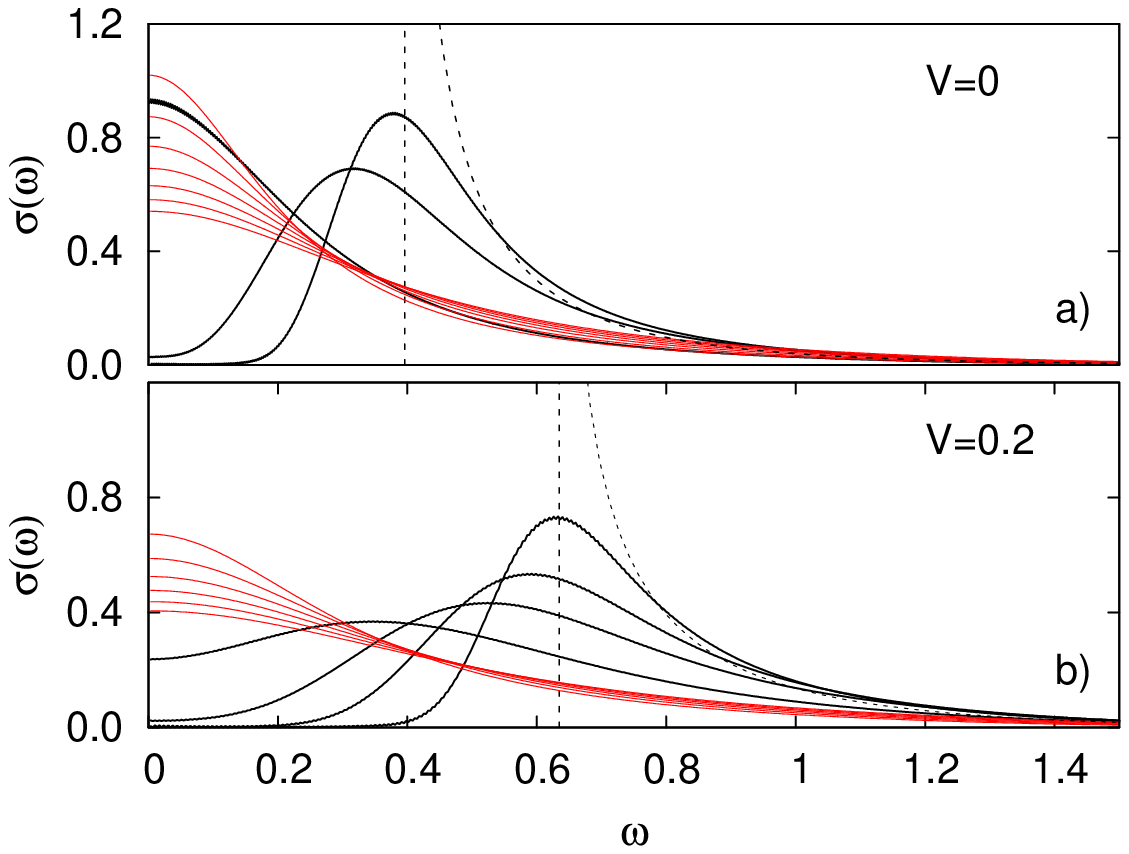}
\caption{(color online) Optical conductivity at  $\lambda=0.4$, $V=0$ (a) and
  $V=0.2D$ (b). The curves in panel (a) 
  are at equally spaced temperatures in the range $T/T_c=0.4-2.9$
  ($T_c=0.026D$) and in  panel (b) $T/T_c=0.2-1.5$ ($T_c=0.052D$).
   Curves corresponding to the disordered
  phase at $T>T_c$ are in red. The
  dashed line is the mean-field result
  Eq. (\ref{eq:hartree_OC}) at the lowest temperature.}   
\label{fig:optcond-wc}
\end{figure}

In figure \ref{fig:optcond-wc} we report the results for the optical
conductivity at $\lambda=0.4$, for different values of
the  intersite Coulomb interaction $V$.   
At such a moderate value  of the  electron-lattice coupling, 
although the overall  behavior qualitatively agrees 
with what is expected in the conventional mean-field
scenario (see Appendix A, and the dashed lines in Fig. \ref{fig:optcond-wc}), 
the spectral features are appreciably smoothened by
the presence of lattice fluctuations. 
A measure of such broadening is provided by 
the variance  $s=\sqrt{2E_P T}$ of the energy fluctuations
of the phonon field,\cite{HoHu,nota-xover}   which is of the order of $0.1 D$
in the example of Fig. \ref{fig:optcond-wc}. 
This is a sizable  fraction of the
electronic bandwidth, which is however sufficiently small that
the overall band picture remains qualitatively valid
at temperatures  comparable or below $T_c$, which are the ones of interest
here.

Due to the classical approximation for the phonons, 
the Drude peak in the normal phase is replaced by an inchoerent gaussian peak 
around $\omega=0$, \cite{millis1} whose width is 
proportional to $s$.
Below the critical temperature, a finite charge disproportionation develops
(cf. Fig. \ref{fig:po}) 
which  is reflected  in the optical spectra  through a
progressive gap opening. 
Correspondingly, the low-frequency spectral weight is transferred 
to frequencies above the optical gap located at 
$\Delta_{opt}\simeq (2 \lambda D +V)\Delta n$.
The sharp square-root divergence at $\omega= \Delta_{opt}$
expected from the standard mean-field treatment
(dashed line, corresponding to 
Eq. \ref{eq:hartree_OC}) is also smoothened due to the presence of 
thermal lattice fluctuations. These  are  also responsible 
for the subgap  absorption tail observed in
Fig. \ref{fig:optcond-wc}, which is absent in the mean-field result. 
As shown in Ref. \onlinecite{Perroni},  a similar
broadening can also arise at $T=0$ due to quantum lattice fluctuations.

The effect of a finite $V$ is to strengthen the CO phase, without modifying the broadening of the spectral features. 
The optical gap increases both due to the
increased value of the charge disproportionation $\Delta n$ (see
Fig.\ref{fig:po}), and to the explicit contribution $V \Delta n$,
which represents the extra electrostatic cost to move one particle from one
sublattice to the other.

\subsection{Strong coupling limit,   $\lambda\gg 1$}

\label{numrestrong}

\begin{figure}[htbp]
\centering
\includegraphics[scale=0.7]{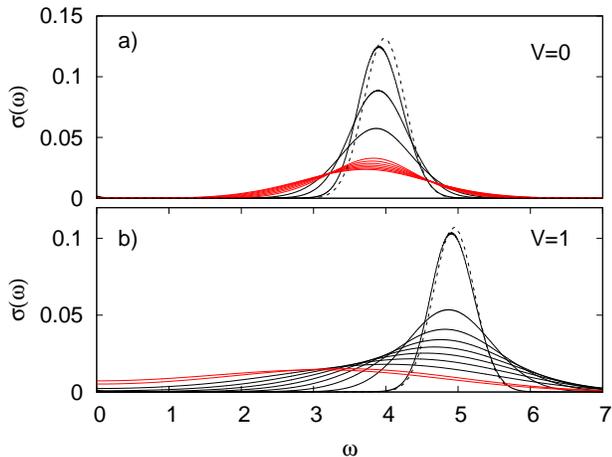}  
\caption{(color online) Optical conductivity at $\lambda=2$, $V=0$ (a) and
  $V=1$ (b). The curves in panel (a) 
  are at equally spaced temperatures in the range $T/T_c=0.3-2.5$
  ($T_c=0.031D$) and in  panel (b) $T/T_c=0.04-1.1$ ($T_c=0.26D$).
 Curves corresponding to the disordered
  phase at $T>T_c$ are in red. The
  dashed line is the strong coupling result
  Eq. (\ref{eq:sigbroken}) at the lowest temperature.}   
\label{fig:optcond-sc}
\end{figure}

The results for the optical absorption in the strong coupling limit 
are shown in Fig.  \ref{fig:optcond-sc} for  $\lambda=2$. 
The  behavior
is very different  from the weak coupling regime analyzed
previously. This is seen already in the normal phase, where the 
Drude peak is replaced  by a broad peak at finite frequency.
This peak arises due
to the formation of small polarons, and reflects the
optical transitions within the polaron internal structure.
The peak position scales with the polaron energy roughly as $\omega \approx
2E_P$, and its width is again 
proportional to the  spread $s$ of the lattice fluctuations.

Because the low-frequency spectral weight is already
strongly suppressed 
in the normal phase,
no clear gap  opening is visible  at $T_c$. 
Indeed, in Fig.
\ref{fig:optcond-sc}.a.  ($\lambda=2$, $V=0$) there is no visible depletion of
the optical conductivity at low frequency  (apart from the natural evolution of
the peak width governed by $s$) and no clear shift of the position of
the polaron peak  as the temperature is reduced below $T_c$. 
Rather, the ordering
transition leads to a sharp increase of the  spectral weight
associated to the polaronic peak.  As we demonstrate hereafter, 
this increase is 
a direct consequence   of the suppression of charge defects,
and constitutes a distinctive signature of the polaronic charge ordering.

The observed behavior can be
understood by noting that in the normal phase, only 
half of the sites neighboring a given polaron are unoccupied, and
therefore available for an optical transition as induced by the
current operator  Eq. (\ref{eq:current}). The number of available
empty neighbors increases as charge defects are progressively removed
below $T_c$,
and so does the weight of the polaronic absorption peak, until each
polaron becomes exclusively surrounded by unoccupied sites at $T=0$.
The above analysis can be carried out in general to show that the
probability for polaronic optical transitions is proportional to
$n_A(1-n_B)$ (for polarons in sublattice A) and
$n_B(1-n_A)$ (for polarons in sublattice B). The sum scales as 
$(1+\Delta n^2)$, leading to an increase of up to a factor of $2$ of
the optical spectral weight in the CO phase.  

A more detailed discussion of the spectral weight enhancement 
is provided in Appendix B, where we derive  analytical expressions
for the optical conductivity valid at large $\lambda$. We
report here the formula appropriate in the CO phase at $V=0$, which
is simply expressed in terms of the normal state polaron absorption
$\sigma_{norm}(\omega)$ as
\begin{eqnarray}
  \label{eq:sigbrokentext}
  \sigma(\omega)&=& 2[1-2n_d(1-n_d)]\sigma_{norm}(\omega).
\end{eqnarray}
This equation 
directly relates the observed enhancement of the polaron peak with the
suppression of the number of defects, in agreement with the
qualitative argument given above, as can be easily shown  using Eq. 
(\ref{eq:defects}) above.
Both this expression and its analog
 Eq. (\ref{eq:sigbrokenV})  valid at
 $V\neq 0$ describe quite accurately  the full DMFT results at $\lambda=2$, 
as illustrated in Fig. \ref{fig:optcond-sc}.

\subsection{Spectral weight analysis}

\begin{figure}[htbp]
\centering
\includegraphics[scale=0.65]{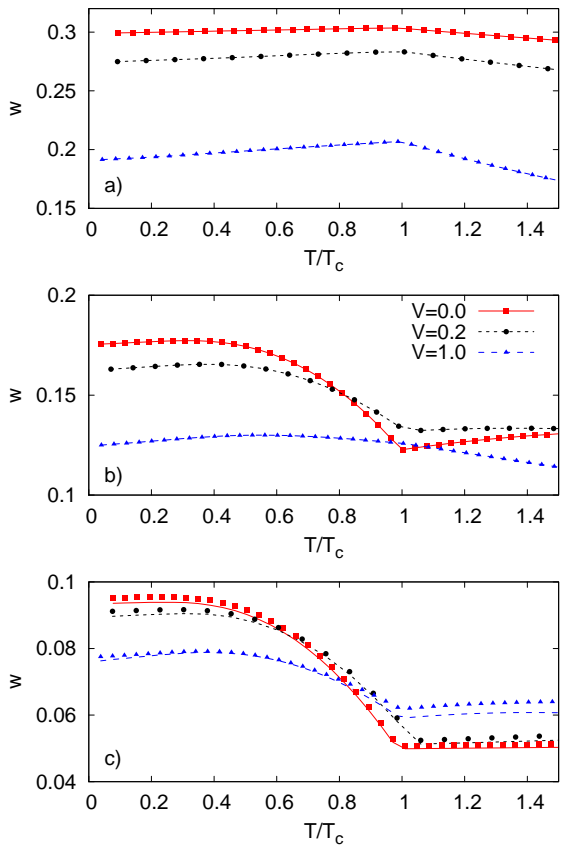}  
\caption{Total optical weight $W$ as defined in Eq (\ref{eq:optcond_sw}), for 
$\lambda=0.4$ (a), $\lambda=1$ (b) 
 and  $\lambda=2$ (c), at different values of $V$ (the legend is 
indicated in panel b).  
Lines in panels (a) and (b) are guides to the eye,
while in panel (c) they correspond to the analytical 
strong coupling approximation  obtained by integrating  
Eqs. (\ref{eq:sigbroken}) and (\ref{eq:sigbrokenV}) down to a cutoff frequency equal to $T$.}
\label{fig:optcond_sw}
\end{figure}

The increase of spectral weight associated to the  ordering of polarons
is best visualized by analyzing the integrated 
optical conductivity $W$ defined in Eq. (\ref{eq:optcond_sw}). 
The evolution of  $W$ {\it vs.} temperature 
is reported in  Fig. \ref{fig:optcond_sw}, and exhibits a markedly
different behavior at weak and strong electron-phonon coupling.
In Fig. \ref{fig:optcond_sw}.c, corresponding to $\lambda=2$ and $V=0$,
we see that upon lowering the temperature 
$W$ first decreases (in the normal phase) and then 
exhibits a sharp increase (in the ordered phase), saturating 
at $T\ll T_c$ to a value which is about twice the normal phase
value. Therefore, at large $\lambda$ the critical
temperature  corresponds to a {\it minimum}  of W. This
remains true in all the polaronic regime, down to the polaron crossover at 
$\lambda\sim 1$.
This is opposite to  the usual behavior 
of  charge density waves,
where upon lowering the
temperature the optical weight first increases and then decreases,
reaching a {\it maximum} at the critical temperature. 
Such conventional behavior is recovered at small $\lambda$, as 
illustrated in  Fig. \ref{fig:optcond_sw}.a for
$\lambda=0.4$. 

\begin{figure}
  \centering
  \includegraphics[scale=0.65]{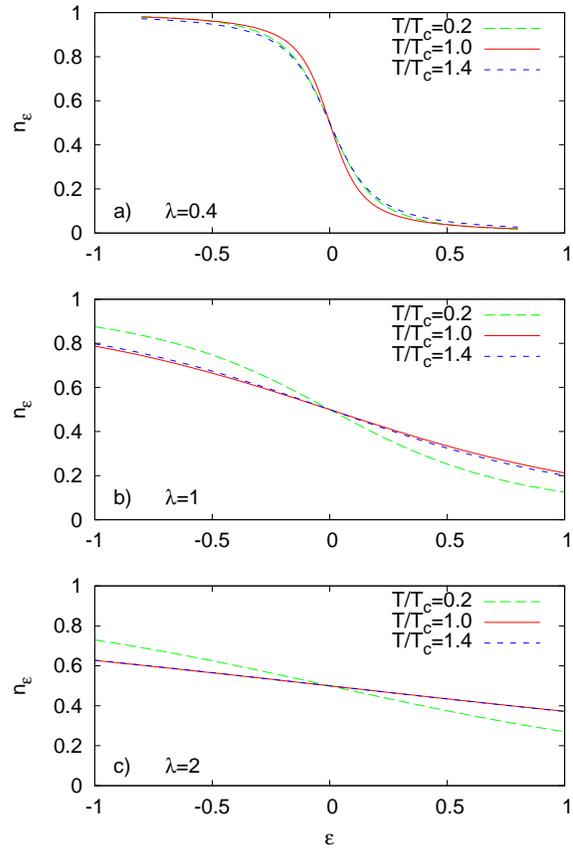} 
  \caption{Momentum distribution function {\it vs.} band energy
    $\epsilon=\epsilon_k$ at temperatures at, below and above $T_c$.
    The coupling strengths are the same as in Fig.\ref{fig:optcond_sw}, and
    we have set $V=0$.} 
  \label{fig:nk}
\end{figure}
These opposite behaviors can be understood in
virtue of the optical sum rule demonstrated in Appendix C, which
relates the total spectral weight to  (minus) the total 
kinetic energy $K$ of the system. The latter is usually
expressed as $K=\sum_k \epsilon_k n_{\epsilon_k}$, where 
$n_{\epsilon_k}$ is the momentum distribution function of the
interacting electrons. As shown in Appendix C,
an equivalent expression holds true even 
in the broken-symmetry phase [see Eq.  
(\ref{eq:totKE})], if one properly introduces the 
occupation number for the generalized propagator $\bar G=
1/[\sqrt{z_Az_B}-\epsilon_k]$. Plots of this quantity are shown in
 Fig.\ref{fig:nk}.

As pointed out in Ref. \onlinecite{Marsiglio}, 
any broadening of the step feature in $n_{\epsilon_k}$ around the
Fermi energy, by populating states with higher energy, leads to an
increase of $K$, and therefore to a decrease of the spectral weight
$W$.  In the case of a conventional charge density wave, 
illustrated in Fig.\ref{fig:nk}.a,
the distribution $n_{\epsilon_k}$ broadens both upon 
increasing the temperature above $T_c$ and upon
opening a gap below $T_c$. As a result, the critical temperature
identifies a minimum of $K$, i.e. a maximum of $W$. 

%Conversely, 
The situation is different in the strong electron-phonon coupling
regime (Fig. \ref{fig:nk}.b and c), where the formation of small
(local) polarons involves all the states in the Brillouin zone.
Correspondingly, the distribution $n_{\epsilon_k}$  becomes 
extremely flat. In this case, increasing the temperature above $T_c$
progressively destroys the local correlations that build up the 
polaron, and the momentum distribution eventually recovers some
structure at the Fermi level,
% which eventually restores some structure in the momentum distribution,
%momentum-space structure and 
%eventually some structure in the momentum distribution is restored, 
%the momentum distribution eventually 
%acquires more structure,
%recovers some structure at the Fermi level, 
leading to a decrease of $K$ (the effect is small because $T_c$ in this
regime is much smaller than the energy scale $E_P$ that governs the
polaron dissociation). The polaron ordering below $T_c$ also decreases
$K$, because 
%the momentum distribution 
the emergence of two distinct sublattices effectively 
%provides 
restores some momentum-space structure.
%effectively restores some structure in momentum space. 
In the
polaronic regime, the critical temperature therefore defines a  maximum of $K$,
i.e. a minimum of $W$.

\subsection{Polaronic regime, $\lambda\gtrsim 1$}

By following the evolution of the spectral weight 
from Fig. \ref{fig:optcond_sw}.a to c,
it is interesting to see that 
at intermediate values of the 
electron-lattice coupling (but still in the polaronic regime), 
the two competing trends evidenced above
lead to a non-monotonic temperature dependence of the kinetic energy
(and of the spectral weight)
within the CO phase. This can be seen  in particular the curve labeled $V=0$ at
$\lambda=1$ in  Fig. \ref{fig:optcond_sw}.b.  
Note that a non-monotonic behavior of the kinetic energy
analogous to the one described here 
was recently observed in the ordered phase of
the Falicov-Kimball model at intermediate $U$.\cite{Matveev08}

\begin{figure}[htpb]
\centering
\includegraphics[scale=0.7]{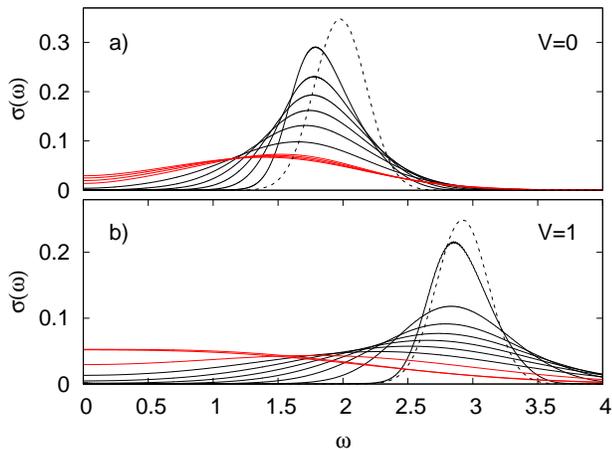}  
\caption{(color online) Optical conductivity at $\lambda=1$, $V=0$ (a) and
  $V=1.0$ (b). The curves in panel (a) 
  are at equally spaced temperatures in the range $T/T_c=0.2-1.5$
  ($T_c=0.052D$) and in  panel (b) $T/T_c=0.04-1.1$ ($T_c=0.25D$).
 Curves corresponding to the disordered
  phase at $T>T_c$ are in red. The
  dashed line is the strong coupling result
  Eq. (\ref{eq:sigbroken}) at the lowest temperature.}   
\label{fig:optcond-ic}
\end{figure}

The  spectral weight analysis presented above suggests that  both features
characteristic of the
weak- and strong-coupling limits coexist in the polaronic
regime at intermediate coupling strengths $\lambda \gtrsim 1$. 
We can check  in Fig. \ref{fig:optcond-ic}.a 
($\lambda=1$, $V=0$) that such coexistence also naturally manifests
in the optical spectra. Here, as in the strong-coupling limit of
Fig. \ref{fig:optcond-sc}.a, the polaronic
nature of the carriers is testified by the presence of a broad
finite-frequency peak already in the normal phase. 
In this case, however, where the polaron energy $E_P$ 
is comparable with the  free electronic bandwidth $D$,  the peak
position lies well below the strong-coupling estimate $2E_P$.
Also the shape and width of the peak deviate from the strong-coupling
estimate, being both controlled by the ratio $s/D$ 
(the usual symmetric gaussian shape is recovered as $s\gg D$, see
Ref. \onlinecite{PRBoptcond}).
Most interesting however is the evolution of 
 the optical absorption 
 below $T_c$, which 
shares similarities with both behaviors shown in
Figs. \ref{fig:optcond-wc}.a and \ref{fig:optcond-sc}.a. 
Indeed, at intermediate values of the
electron-phonon coupling, 
the optical absorption  in the ordered phase exhibits both a marked 
enhancement and  an appreciable transfer of 
 spectral weight  to higher frequencies. This composite behavior gives rise
to an ``isosbestic'' region at frequencies below the polaron peak, 
where the optical absorption is almost independent of temperature.

It was observed  by analyzing the phase diagram in Sec. II that 
 the intersite repulsion $V$ effectively pushes the system 
towards the weak electron-lattice coupling regime.
This conclusion
is also supported 
by the behavior of the optical conductivity spectra. 
For example,  we see from Fig. \ref{fig:Tc} that 
at $\lambda=1$, a repulsion  $V=1$ is sufficient to move 
 the system  outside the polaronic
phase. Accordingly, the polaron peak that was present at high temperature 
in Fig. \ref{fig:optcond-ic}.a has disappeared in the $V=1$ spectrum of 
Fig. \ref{fig:optcond-ic}.b, 
and a more conventional gap opening is restored below $T_c$.
If instead one starts from the strong
coupling value $\lambda=2$,
an appreciable transfer spectral of spectral weight below the
transition is  recovered for  $V=1$, as well as a non monotonic
behavior of the total spectral weight $W$ (see Fig. \ref{fig:optcond-sc}.b and
Fig.\ref{fig:optcond_sw}.c).

\section{d.c. conductivity}
\label{transport}

\begin{figure}
  \centering
\includegraphics[width=80mm]{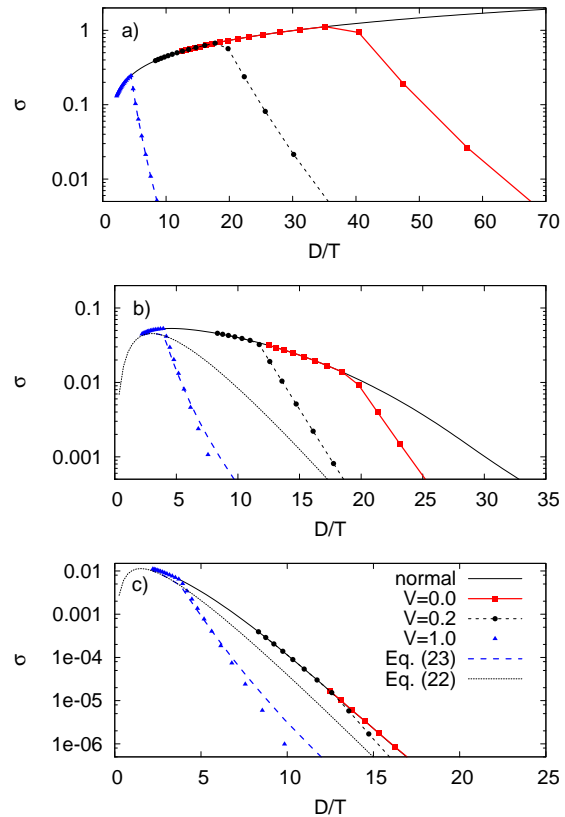}    
  \caption{d.c. conductivity for $\lambda=0.4$ (a), $\lambda=1$ (b) and 
  $\lambda=2$ (c), for different values of $V$.  
 Lines are guides to the eye, except for $V=1$ in
  panels (b) and (c), where the line is the strong coupling formula
  Eq. (\ref{eq:sigdcstrong}). The
  full black line is the DMFT result obtained in
  the normal phase, and the thin dashed line is the standard
  polaronic expression Eq. (\ref{eq:sigdcnormtext}). 
 \label{fig:dc}}
\end{figure}

In this section we briefly describe 
the d.c. conductivity, which is 
obtained from the previously derived Kubo formula Eq. (\ref{OC}), 
by taking the limit $\omega\to 0$.
The numerical results for $\sigma_{d.c.}$ 
across the CO transition are shown
in Fig. \ref{fig:dc} for different values of $V$, at 
$\lambda=0.4$ (a), $\lambda=1$ (b) and $\lambda=2$
(c). These values are the same as analyzed 
in the preceding Section,
and correspond respectively to the weak, intermediate and to the strong
electron-lattice coupling regime.
The CO transition
can be clearly identified as a knee in all the conductivity
curves, except for $\lambda=2,V=0$ (in that case $T_c$ falls outside the range
of Fig. \ref{fig:dc}.c)
In the weak coupling regime (Fig. \ref{fig:dc}.a),  
the opening of a gap changes the 
mobility from metallic-like to insulating-like, i.e. the slope of
$\sigma_{d.c.}(T)$ changes sign.   
The changes in the transport
properties at the CO transition become less marked in
the polaronic regime. 
In this case the mobility is already 
thermally activated in the normal phase, and the CO
transition only  causes an increase of the activation barrier, which
is reflected 
in a change of  slope in the Arrhenius plots
 of  Figs. \ref{fig:dc}.b and Fig. \ref{fig:dc}.c.

An expression for the d.c. conductivity valid in the strong coupling
limit is derived in Appendix B. In the normal phase, one recovers
the usual formula for the polaronic activated behavior: 
\begin{equation}
  \label{eq:sigdcnormtext}
\sigma^{norm}_{d.c.}(T)= 
\frac{\sigma_0}{32 \sqrt{2\pi E_P }T^{3/2}}   e^{- E_P/2T}.
\end{equation}
At  temperatures below $T_c$, this generalizes to
\begin{equation}
  \label{eq:sigdcstrong}
\sigma_{d.c.}(T) = 4 n_d (1-n_d)  \sigma^{norm}_{d.c.}(T).
\end{equation}
We see that the ordering transition reduces the conductivity in a way
that is once again related to the suppression of defects, being
proportional to the product of the occupations in the two
sublattices, namely $4 n_d (1-n_d)\simeq 1-\Delta n^2$.

It should be stressed that standard approaches that incorporate 
Coulomb interactions into the polaronic conductivity
would predict an additive contribution $V\Delta n/2$ to the activation
barrier at the CO transition,\cite{Ihle} which
naturally arises from the extra energy cost that a polaron in the majority
sublattice has to overcome to hop to a neighboring unoccupied site. 
While such contribution is indeed 
recovered in our treatment  [see Eq. (\ref{eq:sigdclo}) in Appendix
B], we see here that the
dominant conduction mechanism at finite temperatures 
is  related to the presence of charge
defects.  Being already thermally excited onto the minority sublattice, 
such defects 
do not pay an extra cost in Coulomb energy when hopping from site to
site, and their motion is therefore more advantageous than ordinary
hopping across the gap
 (remind that at large $\lambda$ the number of defects $n_d$
does not depend explicitly on $V$, as was shown in Fig. \ref{fig:po}).

Eq. (\ref{eq:sigdcstrong}) is compared with the 
full DMFT results for $\lambda=1,V=1$
and $\lambda=2,V=1$ in Fig. \ref{fig:dc}.b and Fig. \ref{fig:dc}.c
respectively.  
It should be stressed that the strong coupling analysis underlying
Eqs. (\ref{eq:sigdcnormtext}) and
(\ref{eq:sigdcstrong}) above is strictly valid only at $\lambda\to
\infty$, as it
 neglects corrections proportional to $D$ in the
activation barrier. As a result, the above formulas widely underestimate the
electrical conductivity,\cite{PRLrhopolaron} and the agreement with
the DMFT data at finite $\lambda$ and $V=0$
is not as good as the one obtained for $\sigma(\omega)$
in the preceding Section. This is particularly evident when comparing
Eq. (\ref{eq:sigdcnormtext}) with the $\lambda=1$ data of
Fig. \ref{fig:dc}.b, where large deviations arise especially at low
temperatures $T\ll E_P$.
Increasing $V$ shifts the ordering transition to
higher temperatures, apparently improving the accuracy of
the approximation.

\section{Discussion and conclusions}

We have calculated the optical and electrical transport properties
in a model describing charge ordering in 
systems with strong electron-lattice interactions. The DMFT treatment
used here properly accounts for fluctuations of the lattice polarization. 
This allows to understand the role played by charge defects 
in the electrodynamic response of the polaronic ordered phase. The effects of 
a moderate intersite Coulomb repulsion term, which is often present in
interacting narrow-band systems, have also been incorporated.

In the weak  electron-phonon
coupling regime, we have seen in Sec. II 
that lattice fluctuations play an important role 
as soon as $\lambda>0$, as they
strongly reduce the critical
temperature compared to the mean-field BCS-like 
expectation.\cite{Ciuk99,Blawid} 
Our results show that these fluctuations also manifest in the
spectral properties, as they  provide  
a disordered effective medium that scatters the electron motion, 
leading to  an appreciable broadening of
the sharp spectral features predicted by mean-field theory.

The effects of lattice fluctuations are however 
much more dramatic in  the strong coupling regime, where the charge
ordering transition occurs in a polaronic phase with
a preexisting charge localization. In that case, 
polaron formation moves most of the spectral weight to a finite frequency
absorption peak, whose position hardly changes at the ordering transition.
Instead, the distinctive
signature of charge  ordering in the strong coupling limit is
a marked enhancement of the polaronic absorption peak,  that is
directly related to the suppression of charge defects, i.e. charges
promoted to the ``wrong'' sublattice by the thermal fluctuations. 
In the polaronic ordered phase, such defects also govern the transport
properties, as their motion   dominates over the usual hopping mechanism
across the disproportionation gap. The competing scenarios described above
appear to coexist 
at intermediate electron-lattice interactions ($\lambda \gtrsim 1$), 
i.e. precisely in the region of the phase diagram where 
the critical temperature is maximum. In this region,  
polaronic charge ordering reveals in the optical spectra 
through both a transfer of spectral weight to high frequencies and a global 
spectral weight enhancement as the temperature is lowered below $T_c$.

It should be noted that a
model very similar to the one studied here was developed in a series
of papers by Ihle and Lorenz,\cite{Ihle,IhleLorenz1,IhleLorenz2} and was
applied to the 
long-standing problem of  the Verwey transition 
in Fe$_3$O$_4$ --- a prototypical example of polaron ordering (see
also the review paper Ref. \onlinecite{Walz}). In
these works the model Eq. (1) was solved relying on the
anti-adiabatic approximation for  the lattice degrees of freedom,
where the phonon dynamics is assumed to be faster than the electrons. 
Our adiabatic treatment
can therefore be considered 
as complementary to the one of Refs. \onlinecite{IhleLorenz1,IhleLorenz2}.
While the present approach clearly gives more insight into the physics of
charge defects, our classical treatment of phonons has an important
limitation that one should keep in mind when attempting a
comparison with actual experiments. The results
obtained here are expected to apply at temperatures and frequencies larger
than the frequency scale of the phonons involved in 
the CO transition. 
Properly accounting for the quantum nature of the phonons would
restore the possibility of  a coherent transport
regime at low temperatures,\cite{PRLrhopolaron} and allow 
for the presence of charge defects even at zero temperature.\cite{Blawid2}

\section*{Acknowledgements.}

The authors acknowledge useful discussions with L. Benfatto,
E. Cappelluti, M. Dressel,  G. Sangiovanni and A. Toschi. S. F. is
grateful for the kind hospitality of the Instituto de Ciencia de
Materiales de Madrid, CSIC, where part of this work was done. 

\appendix
\section{Mean-field treatment of the optical conductivity.}

We evaluate here the optical absorption in the CO state at
mean-field level, which applies to the weak electron-lattice coupling
limit. 
This treatment
 also describes accurately the numerical data at any finite $V$ at nonzero
temperature when 
$\lambda=0$, due to the intrinsic mean-field treatment of
non-local interactions in DMFT. 
The mean-field solution for the electron 
propagators is obtained by inverting the matrix
\begin{equation} \label{eq:invert_V}
\hat{G}_{\bf k}^{-1}=
\left(
\begin{array}{cc}
\omega +\Delta & -\epsilon_k \\
-\epsilon_k &\omega-\Delta 
\end{array}
\right)
\end{equation}
where $\Delta$ is the CO gap, which is determined
self-consistently.\cite{Ciuk99} It is related to the charge
disproportionation by $\Delta=(\lambda D+V/2) \Delta n$.

The corresponding spectral functions are delta functions, which allow
for a direct evaluation of Eq. (\ref{OC}) for a generic DOS. The 
finite-frequency part, corresponding to interband transitions (across
the gap) reads 
\begin{equation}
  \label{eq:hartree_OC}
  \sigma(\omega)= \frac{\sigma_0}{2} \frac{N \phi \left(\frac{1}{2} 
\sqrt{\omega^2-(2\Delta)^2} \right)}
{\sqrt{\omega^2 -(2\Delta)^2}}  
\frac{(2\Delta)^2}{\omega^2}
\tanh    \left(\frac{\omega}{4 T}\right)
\end{equation}
which has a square-root singularity at $\omega=2\Delta$ (the optical
gap) followed by a power-law decay at higher frequencies. At finite
temperature, states are thermally excited across the gap, which
enables the possibility of intraband absorption. In the absence of
additional inelastic scattering mechanisms, this gives rise to a zero-frequency
peak:
\begin{equation}
  \label{eq:Hartree_delta}
  \sigma_D(\omega)=\sigma_0\delta(\omega) \int_{RBZ} d\epsilon 
N(\epsilon)\phi(\epsilon)
\frac{ 2  \epsilon^2}{\epsilon^2+\Delta^2} \left[ 
-f^\prime (\epsilon^2+\Delta^2)\right].
\end{equation}

\section{Strong coupling approximation to $\sigma(\omega)$}

We derive here analytical
approximations to  Eqs. (\ref{eq:self-con}), (\ref{eq:sigma-bub}) 
and (\ref{eq:optcond_sw}) that are valid in the strong coupling limit
$\lambda \gg 1$. 

\subsection{Normal state}

Let us first note that the usual strong-coupling formula of
Reik\cite{Reik} for the optical confuctivity is recovered in the normal phase 
if we take the atomic limit ($t=0$) in the 
propagators appearing in Eq. (\ref{OC}).
Indeed, by setting  $G_0^{-1}=\omega+i \delta$ it can be shown that 
the lattice PDF  is a sum of two gaussians
centered at $\pm g/2k$, each one carrying a weight
$n_A=n_B=1/2$ \cite{Ciuk99} (as was mentioned above, the Coulomb
interaction term does not affect the properties of the normal phase at
mean-field level). This can be cast in the form
\begin{equation}
P_{norm}(X)=g\; \frac{\cosh(\beta  g X /2)}{\sqrt{4 \pi E_P T}}  
 \; 
e^{-\displaystyle\frac{(gX)^2+E_P^2}{4E_P T}}  
 \label{eq:PXat}
\end{equation}
with the polaron energy defined as $E_P=g^2/2k = \lambda D$.
The spectral function is obtained
straightforwardly from Eq. (\ref{eq:PXat}) through the relation
$\rho_{norm}(\omega)=P_{norm}(\omega/g)/g$ 
[cf. Eq. (\ref{eq:self-con})]. 

Since in this limit the electron Green's function is site diagonal and
momentum-independent,  the $\epsilon$-integration can be
factored out from Eq. (\ref{OC}), yielding a prefactor $\int d\epsilon
N(\epsilon)\phi(\epsilon)=1/4$ on the Bethe lattice. 
The remaining frequency integral can be performed using the relation
\begin{equation}
  \label{eq:relation}
  [f(\omega+\nu)-f(\nu)]=\frac{\sinh[\frac{\beta \omega}{2}]}{2 \cosh[\frac{\beta(\omega+\nu)}{2}]
\cosh[\frac{\beta \nu}{2}]}
\end{equation}
which leads to the desired result
\begin{equation}
  \label{eq:signorm}
  \sigma(\omega)=\frac{\sigma_0}{\sqrt{2\pi E_P
      T}}\frac{\sinh(\beta \omega/2)}{16\omega} 
e^{-\displaystyle \frac{ \omega^2+4E_P^2}{8E_P T}}.
\end{equation}
Focusing on the finite frequency part $\omega\gg T$,
this can be further simplified to 
\begin{equation}
  \label{eq:sigreik}
  \sigma_{norm}(\omega)=\frac{\sigma_0}{32\omega \sqrt{2\pi E_P
      T}} 
e^{-\displaystyle \frac{ (\omega-2E_P)^2}{8E_P T}}.
\end{equation}
This formula represents a modified gaussian absorption band having its
maximum at 
$\omega=2E_P$ and a width $\sqrt{2}s=2 \sqrt{E_P T}$, which 
reflects the thermal 
fluctuations of the phonon field that couples to the electron motion. 
Note that if the phonon quantum
fluctuations are correctly taken into account, the width of the
absorption band does not shrink indefinitely at low temperatures, but
rather saturates to a finite value $s^2=2 E_P \omega_0$.\cite{PRBoptcond} 
One can observe that the numerical factor $1/32$ in Eq. (\ref{eq:sigreik}) 
implicitly includes  the density dependent factor $n(1-n)$ expected from 
theories   of independent polarons.

\subsection{Charge ordered state}

\label{gaussian}

In the charge ordered state ($\Delta n\neq 0$), 
the two contributions to the lattice PDF are
unbalanced by a factor which takes into account the different fillings
$n_A\neq n_B$ on nonequivalent sublattices. As a result, Eq. (\ref{eq:PXat})
becomes
\begin{equation}
  \label{eq:PXco}
  P_{A,B}(X)=P_{norm}(X) [1\pm  (1-2n_d) \tanh(\beta g X/2)]
\end{equation}
where the $+$ and $-$ signs correspond to the $A$ and $B$  sublattices
respectively. The lattice PDF is again the sum of two gaussian
peaks, centered  at the minima $\pm X_0$ of the adiabatic 
potentials $V_{A,B}$, 
whose weights are now respectively $n_d$ and $1-n_d$.
At $\lambda=2$, this approximation compares very well 
with the full DMFT result illustrated in Fig. \ref{fig:defects}. 
That figure also shows
that for moderate $V\lesssim 1$
the shape of the gaussian peaks in the PDF is unaffected by the electrostatic
repulsion, 
which only enters implicitly through the self-consistent
determination of $n_d$. 

From Eq. (\ref{eq:PXco}) we obtain the spectral function for the two
sublattices: 
\begin{eqnarray}
\rho_A(\omega)&=&(1-n_d) \ g_+\left(\omega+\frac{V\Delta n}{2}\right)+
n_d \ g_-\left(\omega+\frac{V\Delta n}{2}\right)\nonumber \\
\rho_B(\omega)&=&n_d \ g_+\left(\omega-\frac{V\Delta n}{2}\right)+
(1-n_d) \ g_-\left(\omega-\frac{V\Delta n}{2}\right)\nonumber
\end{eqnarray}
where
\begin{equation}
g_{\alpha}(\omega)=
\frac{1}{\sqrt{2\pi s^2}}
\exp\left(-\frac{\omega-\alpha E_P}{2s^2}\right)
\end{equation}
and $\alpha=\pm 1$.

Taking the low temperature limit ($T\ll \sqrt{E_P T},D$) in
Eq. (\ref{OC}) and using the spectral functions above, 
it is possible to obtain an expression for the
finite frequency conductivity  ($\omega \gg T$). At $V=0$ it takes the
simple form:
\begin{eqnarray}
  \label{eq:sigbroken}
  \sigma(\omega)&=& 2[1-2n_d(1-n_d)]\sigma_{norm}(\omega)
\end{eqnarray}
which directly relates the observed enhancement of the
optical absorption in the CO phase, in agreement 
 with the general arguments presented in Sec. \ref{numrestrong}.
In the presence of intersite repulsion, the above formula generalizes
to 
\begin{eqnarray}
  \label{eq:sigbrokenV}
  \sigma(\omega)&=&\frac{\sigma_0}{16\omega \sqrt{2\pi E_P T}}
\left[
e^{-\displaystyle \frac{ (\omega-2E_P-V\Delta n)^2}{8E_P T}}
(1-n_d)^2 \right. \nonumber \\
 &+& \left. e^{-\displaystyle \frac{ (\omega-2E_P+V\Delta n)^2}{8E_P T}}
(n_d)^2
\right], 
\end{eqnarray}
which clearly reduces to Eq. (\ref{eq:sigbroken}) when $V=0$.
Both  expressions  compare well with  the numerical data of
Fig. \ref{fig:optcond-sc} at $\lambda=2$. Moreover,
the total optical spectral weight obtained by integrating
Eq. (\ref{eq:sigbrokenV}) down to a cutoff frequency $\omega=T$ 
correctly reproduces the non-monotonic temperature dependence of
the DMFT results shown in Fig. \ref{fig:optcond_sw}.c.
Note that 
at extremely large $V$, this expression  predicts a double-peak
structure, which is not observed in the numerical data at moderate $V$.

The  results presented in this Appendix,
derived under the assumption of a vanishing bandwidth, 
are strictly valid as long as
$s\gg D$. \cite{PRBoptcond}  From the asymptotic strong coupling
expression  $T_c=D/(16 \lambda)+V/4$ 
 we can estimate
\begin{equation}
  \label{eq:estimate}
  \left.\frac{s}{D}\right \vert_{T_c}\simeq\frac{1}{2^{3/2}} 
\sqrt{1 +4\lambda\frac{V}{D}}.
\end{equation}
We see that  at $V=0$
the condition $s\gg D$   
is never  realized in the relevant region around $T_c$, and some
finite bandwidth corrections to the polaronic lineshapes can be expected. 
\cite{PRBoptcond}   Examination of Figs. \ref{fig:optcond-sc} and 
\ref{fig:optcond-ic} shows that  the accuracy of the strong
coupling formula Eq. (\ref{eq:sigbroken})  improves when
an explicit Coulomb term $V$ is included.

\subsection{d.c. conductivity}

The d.c. conductivity is obtained by taking the limit  $\omega\to 0$
in Eq. (\ref{OC}). Some care must be taken in
integrating the gaussian spectral functions in the presence of the
factor  $1/[4T\cosh  (\nu/2T)]$ originating from the  
Fermi functions. For $T\ll E_P$ one obtains  
\begin{eqnarray}
  \label{eq:sigdc}
  & & \sigma_{d.c.}(T)=\frac{\sigma_0}{16\pi E_P T} e^{-\frac{E_P}{2T}-\frac{(V
      \Delta n)^2}{8 E_P T}} \times \\ \nonumber
& \times & \left[ 2n_d(1-n_d) \sqrt{\frac{\pi
      E_P}{2T}} +(1-n_d)^2 e^{-\frac{V\Delta n}{2T}} + n_d^2
e^{\frac{V\Delta n}{2T}}
\right].
\end{eqnarray}
which reduces in the normal phase to the usual formula for the polaronic
conductivity:   
\begin{equation}
  \label{eq:sigdcnorm}
\sigma^{norm}_{d.c.}(T)= 
\frac{\sigma_0}{32 \sqrt{2\pi E_P }T^{3/2}}   e^{- E_P/2T}.
\end{equation}
In the CO state at temperatures
$T\lesssim T_c$, the leading contribution to Eq. (\ref{eq:sigdc}) 
is
\begin{equation}
  \label{eq:sigdcint}
\sigma_{d.c.}(T) = 4 n_d (1-n_d)   e^{-\frac{(V
    \Delta n)^2}{8 E_P T}} \sigma^{norm}_{d.c.}(T).
\end{equation}
which shows that the dominant conduction mechanism in the polaronic
ordered phase  involves thermally
activated defects.
The exponential term is close to $1$ at moderate values of $V$ such as
the ones studied here, and can be dropped.
This expression breaks down as $T\ll T_c$, when the number
of defects $n_d$ vanishes. In this
case a more conventional result is recovered
\begin{equation}
  \label{eq:sigdclo}
  \sigma_{d.c.}(T) =\frac{\sigma_0}{16\pi E_P T} e^{-\frac{(E_P+V
    \Delta n/2)^2}{2 E_P T}}.
\end{equation}
This formula predicts an additive contribution $V\Delta n/2$ to the
activation barrier, corresponding to the extra energy cost that a
polaron has to overcome when hopping to its neighboring sites.\cite{Ihle}

\section{f-sum rule in the broken-symmetry phase}

In this section we demonstrate the f-sum rule for the optical
conductivity evaluated through the DMFT
formula Eq. (\ref{OC}), generalizing the demonstration of Ref.
\onlinecite{Blumer} to  the broken-symmetry phase. 
We first rewrite the integral of the optical conductivity in the form
\begin{equation}
  \label{eq:intOC}
  \int_0^\infty \sigma(\omega)=\frac{\sigma_0}{2} \int_{RBZ} d\epsilon
  N(\epsilon)\phi(\epsilon) I(\epsilon),
\end{equation}
with 
\begin{widetext}
\begin{eqnarray}
  \label{eq:Ieps}
  I(\epsilon)&=&\frac{1}{\pi^2}\int_{-\infty}^{\infty}d \nu 
\int_{-\infty}^{\infty} d \nu^{\prime} 
\frac{f(\nu)-f(\nu^\prime)}
{\nu^\prime-\nu} Tr \left[
\sigma_x Im \hat{G}_\epsilon(\nu^\prime)\sigma_x Im \hat{G}_\epsilon(\nu)
\right].
\end{eqnarray}
\end{widetext}
This expression can be transformed to
\begin{eqnarray}
I(\epsilon)&=& \frac{1}{\pi}\int_{-\infty}^{\infty}d \nu f(\nu)
Tr \left[2
\sigma_x Im \hat{G}_\epsilon(\nu)\sigma_x Re \hat{G}_\epsilon(\nu)
\right]\nonumber \\
&=& \frac{Tr}{\pi}\int_{-\infty}^{\infty}d \nu f(\nu)
Im  \left[
\sigma_x \hat{G}_\epsilon(\nu)\sigma_x \hat{G}_\epsilon(\nu)
\right]\nonumber\\
&=&
\frac{Tr}{\pi}\int_{-\infty}^{\infty}d \nu f(\nu)
Im  \left[
\sigma_x \frac {d\hat{G}_\epsilon(\nu)}{d\epsilon}
\right]\nonumber.
\end{eqnarray}
The latter equality is obtained using the fact that 
$d(\hat{G}_\epsilon^{-1}\hat{G}_\epsilon)/d\epsilon=d(1)/d\epsilon=0$ 
and observing that from Eq. (\ref{eq:invert}) one has
$d\hat{G}^{-1}_\epsilon/d\epsilon=-\sigma_x$. 
Going back to Eq. (\ref{eq:intOC}) and integrating by parts yields
\begin{equation}
  \label{eq:sumrule}
  \int_0^\infty \sigma(\omega)=\sigma_0 \int_{RBZ}d\epsilon
  \frac{d (N\phi)}{d\epsilon} \int d\nu f(\nu) 
2\rho^{AB}_\epsilon(\nu).
\end{equation}
With the present choice of the current vertex $\phi(\epsilon)$ for the
Bethe lattice one has
$\frac{d N\phi}{d\epsilon}=-\epsilon N(\epsilon)$ so that the above
expression reduces to
\begin{equation}
  \label{eq:sumruleK}
  \int_0^\infty \sigma(\omega)=-\frac{\sigma_0}{2}K
\end{equation}
where $K$ is the total kinetic energy defined as
\begin{equation}
  \label{eq:KE}
  K=\int_{RBZ}d\epsilon 
N(\epsilon)\epsilon \int d\nu f(\nu) 
2\rho^{AB}_\epsilon(\nu).
\end{equation}
Introducing the  following spectral density, $\bar\rho_{\epsilon_k}(\nu)=-Im
[\sqrt{z_Az_B}-\epsilon_k]^{-1}/\pi$, with $z_A$ and $z_B$ defined
after Eq. (\ref{eq:invert}),  and the corresponding
momentum distribution function 
%$n_\epsilon\equiv n_{\epsilon_k}$
\begin{equation}
  \label{eq:nk}
  n_{\epsilon_k}=\int d\nu f(\nu) \bar\rho_{\epsilon_k}(\nu), 
\end{equation}
which is the appropriate generalization of the usual $n_k$
%$n_k=\int d\nu  f(\nu) \rho_{\epsilon_k}(\nu)$ 
to the broken-symmetry phase, the total
kinetic energy Eq. (\ref{eq:KE}) can be rewritten as:
\begin{equation}
  \label{eq:totKE}
K=2 \int_{RBZ}d\epsilon 
N(\epsilon)\epsilon \ n_\epsilon =2\sum_{k \in RBZ} \epsilon_k n_{\epsilon_k}.
\end{equation}


\begin{thebibliography}{99} 
\bibitem{Gruner} G. Gr\"uner, Density Waves in Solids, 
Addison-Wesley Publishing Company, 1994.
\bibitem{Dessau}D. S. Dessau et al., Phys. Rev. Lett. 81, 192 (1998)
\bibitem{Adams} C. P. Adams, J. W. Lynn, Y. M. Mukovskii, A. A. Arsenov, D. A. Shulyatev,  Phys. Rev. Lett. 85, 3954 (2000)
\bibitem{Vasiliu} L. Vasiliu-Doloc, S. Rosenkranz, R. Osborn,
  S.K. Sinha, J.W. Lynn, J. Mesot, O.H. Seeck, G. Preosti, A.J. Fedro,
  and J.F. Mitchell, Phys. Rev. Lett. 83, 4393 (1999) 
\bibitem{Katsufuji}  T. Katsufuji et al., Phys. Rev. B 54, R14230
  (1996).
\bibitem{Calvani} P. Calvani, A. Paolone, P. Dore, S. Lupi, P. Maselli,
P. G. Medaglia, S-W. Cheong, Phys. Rev. B 54, R9592 (1996)

\bibitem{Bernhard} C. Bernhard, A.V. Boris, N. N. Kovaleva,
  G. Khaliullin, A.V. Pimenov, Li Yu, D. P. Chen, C.T. Lin, and
  B. Keimer. Phys. Rev. Lett 93, 167003 (2004)

\bibitem{Wang} N. L.Wang, Dong Wu,  G. Li, X. H. Chen, C. H. Wang, and
  X.G. Luo, Phys. Rev. Lett.  93, 147403 (2004)



\bibitem{Degiorgi} L. Degiorgi, P. Wachter and D. Ihle, Phys. Rev. B
  35, 9259 (1987)

\bibitem{Park} S. K. Park, T. Ishikawa, and Y. Tokura, Phys. Rev. B
  58, 3717 (1998)

\bibitem{Schrupp}D. Schrupp et al.,
Europhys. Lett. 70, 789 (2005)

\bibitem{Presura} C. Presura, M. Popinciuc, P. H. M.  van Loosdrecht,
  D. van der Marel, M. Mostovoy, T. Yamauchi, and Y. Ueda,
  Phys. Rev. Lett. 90, 026402 (2003) 

\bibitem{Baldassarre} L. Baldassarre, A. Perucchi, E. Arcangeletti, D. Nicoletti, D. Di Castro, P. Postorino, V. A. Sidorov, and S. Lupi
Phys. Rev. B 75, 245108 (2007) 

\bibitem{Attfield} J. P. Attfield, A. M. T. Bell,
  L. M. Rodriguez-Martinez, J. M. Greneche, R. J. Cernik, J. F. Clarke
  \& D. A. Perkins, Nature 396, 655 (1998)


\bibitem{Perfetti1} L. Perfetti, H. Berger, A. Reginelli, L. Degiorgi,
  H. Hochst, J. Voit, G. Margaritondo, and M. Grioni,
  Phys. Rev. Lett. 87, 216404 (2001)
\bibitem{Perfetti2}
L. Perfetti, S. Mitrovic, G. Margaritondo, M. Grioni, L. Forro, L. Degiorgi, and H. Hochst, Phys. Rev. B 66, 075107 (2002).

\bibitem{Vuletic} T. Vuletic, B. Korin-Hamzic, T. Ivek, S. Tomic,
  B. P. Gorshunov, M.  Dressel and J. Akimitsu,
Phys. Rep. 428, 169 (2006).

\bibitem{Drichko} M. Dressel and N. Drichko, Chemical Reviews 104, 5689 (2004).



\bibitem{Bulla99} R. Pietig, R. Bulla, S. Blawid,
Phys. Rev. Lett. 82, 4046 (1999) 

\bibitem{Bulla04} N.-H. Tong, S.-Q. Shen, R. Bulla, 
Phys. Rev. B 70, 085118 (2004) 
\bibitem{Daghofer} M. Daghofer, R. M. Noack, P. Horsch, preprint
  arXiv:0711.1990

 
\bibitem{Walz} F. Walz, J. Phys. Cond. Matter 14, R285 (2002)

\bibitem{Ciuk99} S. Ciuchi and F. de Pasquale, Phys. Rev. B 59, 5431 (1999)

\bibitem{Hassan07} S. R. Hassan and H. R. Krishnamurthy, Phys. Rev. B
  76, 205109 (2007)


\bibitem{Matveev08} O. P. Matveev, A. M. Shvaika, J. K. Freericks,
  Phys. Rev. B 77, 035102 (2008) 




\bibitem{Perroni}     C. A. Perroni, V. Cataudella, G. De Filippis, G. Iadonisi, V. M. Ramaglia, and F. Ventriglia, Phys. Rev. B 67, 094302 (2003) 



\bibitem{note-BCS} A BCS-like prediction $T_c\propto \exp(-const/\lambda)$, 
valid in the limit $\lambda\to 0$, actually 
overestimates the critical temperature at any finite $\lambda$ 
 since it neglects the possible 
fluctuations of the order parameter, see Refs. \onlinecite{Ciuk99,Blawid}.


\bibitem{Frat1D} S. Fratini and G. Rastelli, Phys. Rev. B 75, 195103 (2007)

\bibitem{Blawid2}
S. Blawid and A. J. Millis, Phys. Rev. B  63, 115114 (2001)

\bibitem{Blawid} S. Blawid and A. Millis, Phys. Rev. B 62, 2424 (2000)       

\bibitem{Kotliar} A. Georges et al., Rev. Mod. Phys. 68, 13 (1996).
\bibitem{Khurana} A. Khurana, Phys. Rev. Lett. 64, 1990 (1990).
\bibitem{Schweitzer}     H. Schweitzer and G. Czycholl, Phys. Rev. Lett. 67, 3724 (1991)
\bibitem{Jarrell} Th. Pruschke, D. L.  Cox, and M.  Jarrell,  Phys. Rev. B 47, 3553 (1993) 

\bibitem{Freericks} W. Chung and J. K. Freericks, Phys. Rev. B 57,
  11955 (1998). 
\bibitem{Millis} A. Chattopadhyay, A. J. Millis, and S. Das Sarma, Phys. Rev. B
61, 10738 (2000).

\bibitem{PRBoptcond} S. Fratini and S. Ciuchi, Phys. Rev. B 74, 075101
  (2006) 

\bibitem{HoHu} S. Fratini and S. Ciuchi, Phys. Rev. 72, 235107 (2005)

\bibitem{nota-xover} This estimate is valid 
outside the polaronic crossover region. Approaching the polaron crossover
coupling both from above and below the CO temperature, 
the local displacements
strongly deviate from the gaussian behaviour.



\bibitem{millis1} A. J. Millis, R. Mueller and B. I. Shraiman,
  Phys. Rev. B 54, 5389 (1996)

\bibitem{Marsiglio} F. Marsiglio, Phys. Rev. B 73, 064507 (2006)

\bibitem{Ihle} D. Ihle, Z. Phys. B 58, 91 (1985) 

\bibitem{PRLrhopolaron} S. Fratini and S. Ciuchi
Phys. Rev. Lett. 91, 256403 (2003) 

\bibitem{IhleLorenz1} D. Ihle and B. Lorentz, J. Phys. C: Solid State
  Phys. 18, L647 (1985) 
\bibitem{IhleLorenz2} D. Ihle and B. Lorentz, J. Phys. C: Solid State
  Phys. 19, 5239 (1986) 
 
\bibitem{Reik} H. G. Reik, Solid State Commun. 1, 102 (1963)

\bibitem{Blumer} N. Bl\"umer, Ph.D Thesis, Univ. Augsburg (2003) 




\end{thebibliography}
\end{document}